\documentclass{article}
\usepackage{spconf,amsmath,graphicx}

\usepackage{enumitem}
\setlist{nosep, leftmargin=14pt}
\usepackage{mwe} % to get dummy images

\title{A Semi-supervised Learning Approach for B-line Detection in Lung Ultrasound Images}
\name{Tianqi Yang$^{1}$\quad Nantheera Anantrasirichai$^{1}$\quad Oktay~Karakuş$^{2}$\quad Marco Allinovi$^{3}$ \quad Alin~Achim$^{1}$ \thanks{e-mail: tianqi.yang@bristol.ac.uk; n.anantrasirichai@bristol.ac.uk;  karakuso@cardiff.ac.uk; marco.allinovi@gmail.com; alin.achim@bristol.ac.uk}}

\address{$^{1}$ Visual Information Lab, University of Bristol, Bristol, UK\\
        $^{2}$ School of Computer Science and Informatics, Cardiff University, Cardiff, UK\\
        $^{3}$  Nephrology, Dialysis and Transplantation, Careggi University Hospital, Florence, Italy}
\begin{document}
%\ninept
%
\maketitle
\begin{abstract}
Studies have proved that the number of B-lines in lung ultrasound images has a strong statistical link to the amount of extravascular lung water, which is significant for hemodialysis treatment. Manual inspection of B-lines requires experts and is time-consuming, whilst designing automatic methods is currently problematic because of the lack of ground truth.   Therefore, in this paper, we propose a novel semi-supervised learning method for the B-line detection task based on  contrastive learning. Through multi-level unsupervised learning on unlabelled lung ultrasound images, the features of the artefacts are learnt. In the downstream task, we introduce a fine-tuning process on a small number of  labelled images using the EIoU-based loss function. Apart from reducing the data labelling workload, the proposed method shows a superior performance compared to model-based methods with the recall of 91.43\%, the accuracy of 84.21\% and the $F_1$ score of 91.43\%.
\end{abstract}
\section{Introduction}
In recent years, lung ultrasound (LUS) has been increasingly used as a support tool in clinical diagnoses. Being a non-invasive and easy-to-use technique, it is regarded as a prospective routine practice for bedside patient assessment \cite{touw2015lung, touw2019routine}. For hemodialysis patients, an accurate estimation of  extravascular lung water is a primary approach to avoid chronic dehydration or long-term cardiovascular complications \cite{hecking2015consequences}. LUS has been shown reliable for assessing tissue fluid overload \cite{torino2016agreement,vitturi2014lung,panuccio2012chest} through evaluating the volume status of dialysis patients by counting the number of B-lines (B-line scores) \cite{lichtenstein1997comet}. B-line quantification correlates and represents the pulmonary congestion which is representative of the whole body fluid overload. Recent studies \cite{noble2009ultrasound,fu2021lung,allinovi2017lung} have reported that in both adults and children patients the B-line scores are correlated with the volume of extravascular lung water. Therefore, the detection of B-lines becomes essential for excess fluid quantification. 

In lung ultrasonography, B-lines appear as vertical comet-tail artifacts arising from the pleural line. The presence of a few scattered B-lines can be normal, as found in healthy subjects, whilst multiple B-lines are considered the sonographic sign of lung interstitial syndrome \cite{soldati2009sonographic}. This is because the difference in acoustic impedance between the lung and the surrounding tissues is increased when lung density increases due to extravascular fluid \cite{anantrasirichai2016automatic}. 

Following the step of the first reported work in the area of computerized B-lines scoring \cite{brattain2013automated}, various studies have investigated automatic B-line detection methodologies. Anantrasirichai et. al. posed the line detection as an inverse problem \cite{anantrasirichai2016automatic,anantrasirichai2017line}, where a gray scale LUS image is converted to a representation of radius and orientation in the Radon domain. The inverse problem is then solved using the alternating direction method of multipliers. Moshavegh et. al. \cite{moshavegh2018automatic} used a random walk method to delineate the pleural line, and then applied an alternate sequential filtration to identify B-lines in the area that excludes the upper pleural region. In the context of evaluating COVID-19 patients, Karakus et. al. \cite{karakucs2020detection} have improved the line detection performance by regularizing the solution using the Cauchy proximal splitting (CPS) algorithm \cite{karakucs2020convergence}. This promotes statistical sparsity by utilising the Cauchy-based penalty function. However, all these schemes require an initial detection of the pleural line in order to locate the lung space. With the advent of deep learning technique, Van Sloun and Demi \cite{van2019localizing} proposed a weakly-supervised learning method, which exploits 20 convolutional layers to perform B-line localization by gradient-weighted class activation mapping (grad-CAM) \cite{selvaraju2017grad}, but the class activation mapping is far less precise to isolate the B-lines. Baloescu et. al. \cite{Baloescu2020automated} used neural networks with 3-D filters for B-line quantification, but the work was limited to supervised learning. Though Kerdegari et. al. \cite{Kerdegari2021bline} combined the long short-term memory (LSTM) network and temporal attention to achieve B-line localization, the complex structure of LSTM is often demanding on the hardware.

The key challenge of deep learning approaches is the scarcity of ground truth, particularly in medical imaging as creating it is time-consuming and very subjective. Therefore, in this paper, we tackle this problem with a semi-supervised learning framework, comprising i) unsupervised contrastive learning \cite{hadsell2006dimensionality} for B-line feature representation, and ii) transfer learning with a small set of labelled data for B-line detection. The first part of the proposed method is based on DetCo \cite{xie2021detco} technique, where  discriminative representations are jointly learnt from global images and local patches via contrastive learning. A residual network (ResNet) \cite{he2016deep} model is first trained on an unlabelled LUS dataset, where the model learns the features of LUS images. Then the trained ResNet model is employed as a backbone of a Faster-RCNN \cite{ren2015faster} to detect B-lines. The proposed method has two main advantages: i) it detects the B-lines in LUS images directly without the need of pleural line localisation; and ii) being a semi-supervised learning method, it minimizes the annotation workload. To the  best of our knowledge, this is the first time that unsupervised learning is used to inform B-line features in LUS images. This feature extractor consequently improves the performance of B-line detector, where only a small labelled dataset is required for training.
\vspace{-2mm}
% The main purpose of unsupervised learning is to pre-train representations that can be transferred to downstream tasks by fine-tuning \cite{he2020momentum}. 
 % which is one of the biggest challenges in medical image processing. 

%The remaining of this paper is organised as follows. The theoretical background and the method implementation are described in Section \ref{sec:2}. The performance of the proposed method and the comparison with recent method are presented in Section \ref{sec:3}. The discussion in Section \ref{sec:4} interprets the study result and points out future research directions. Finally, Section \ref{sec:5} presents the conclusions of this work.

\begin{figure}[t!]
\label{fig:diagram}
\centering
\includegraphics[width=1\linewidth]{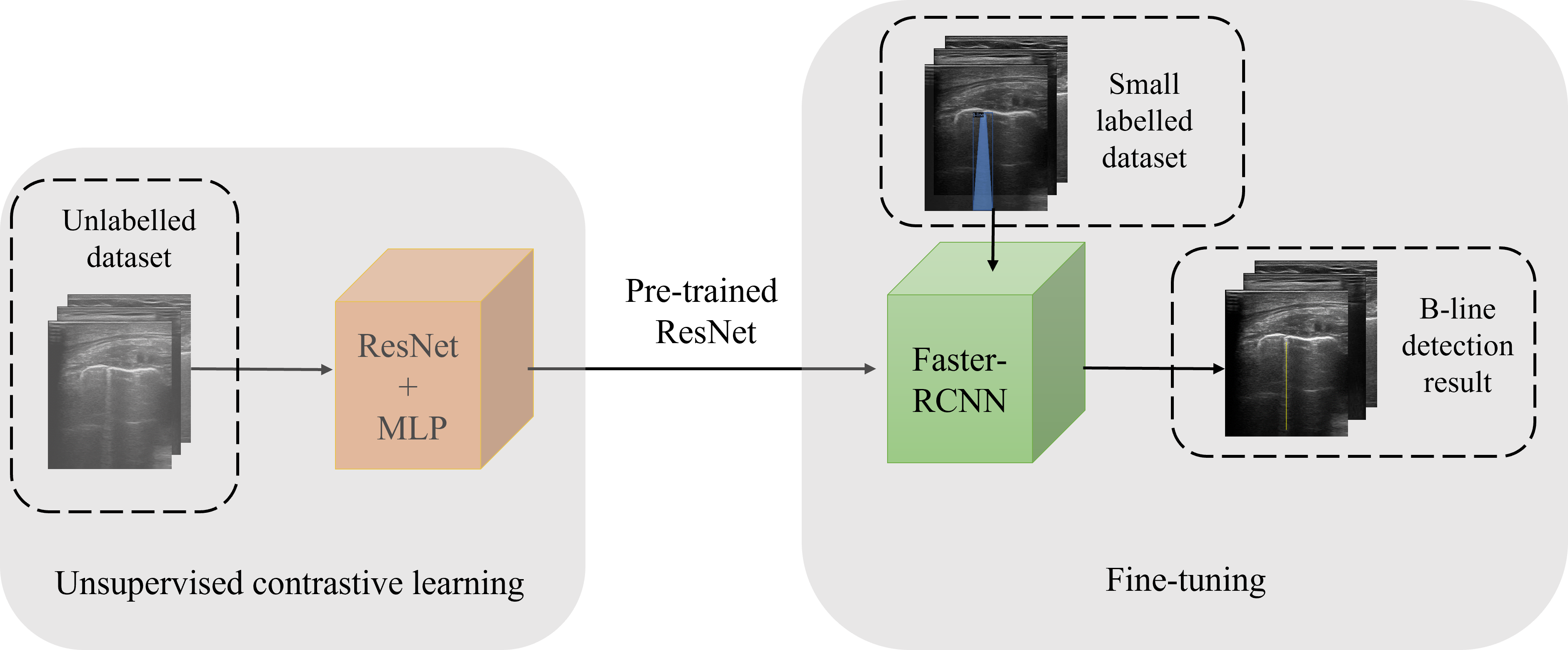}
\caption{The proposed framework. }
\vspace{-5mm}
\end{figure}

\section{Proposed Method}
\label{sec:2}

A diagram of the proposed framework is shown in Fig. \ref{fig:diagram}, comprising B-line feature learning (Section \ref{sec:2.1}) and B-line detector (Section \ref{sec:2.2}).\vspace{-5mm}
\subsection{Feature representation with contrastive learning}\vspace{-2mm}
\label{sec:2.1}

\noindent \textbf{Unsupervised contrastive learning} is one of the most promising directions for the development of unsupervised deep learning \cite{xie2021detco}. The neural network is trained on unlabelled datasets to learn feature representations. This characteristic is valuable in medical image field where labelled data is hardly available. Contrastive learning aims to learn a feature representation in the embedding space where intra-class features have a greater overlap than inter-class features by using an objective function to measure the similarity among the objects. A milestone of using contrastive learning in image processing is MoCo \cite{he2020momentum}. It models the learning process as dictionary look up, where a query is expected to be matched with the positive key and distinguished from other keys. Based on MoCo, Xie et. al. proposed a framework called DetCo \cite{xie2021detco} specifically for object detection. As shown in Fig. 2, the network learns feature representation with multi-level optimization and the framework builds contrastive loss across both the global and local views.

\begin{figure}[t!]
\centering
\includegraphics[width=1\linewidth]{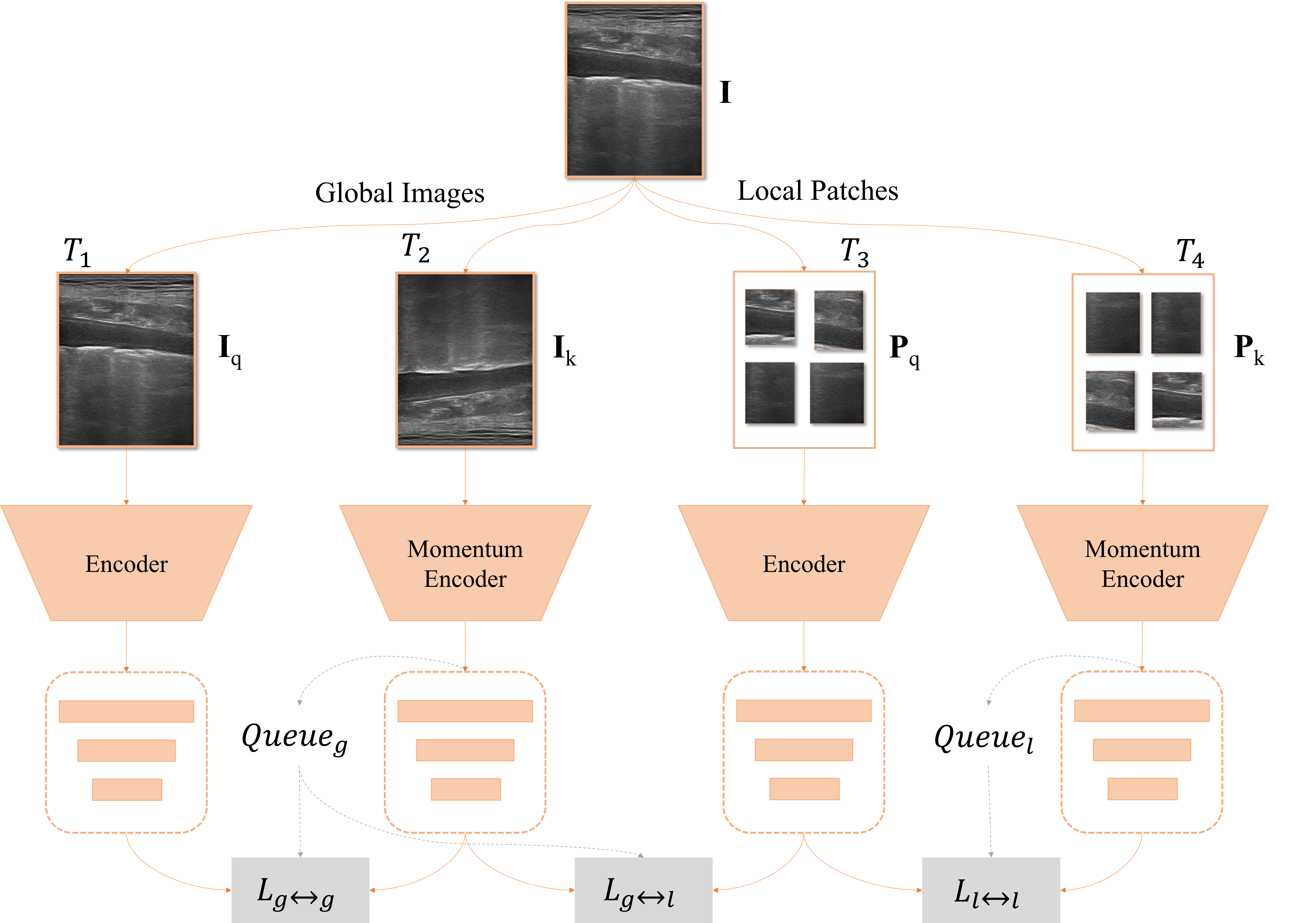}
\caption{The overall pipeline of DetCo \cite{xie2021detco} adapted for LUS. $T$ means image transforms. $Queue_{g/l}$ means memory banks for global/local features, respectively. }
\label{fig:DetCo}
\vspace{-5mm}
\end{figure}

Fig. \ref{fig:DetCo} shows our LUS feature learning based on DetCo. Each image $\boldsymbol{I}$ is transformed into global images and local patches $\boldsymbol{P}$. The encoder and the momentum encoder can be of the same, similar, or different structures. In this paper, we use ResNet \cite{he2016deep} for both encoders and momentum encoders. The former is trained to learn feature representation and output the query, the latter is updated by a momentum parameter and output a queue of keys that are stored in the memory bank \cite{yang2021instance}.

\noindent \textbf{Loss Functions}: The loss used in unsupervised feature learning stage follows DetCo \cite{xie2021detco} and is defined as follows:
\begin{equation}
\begin{aligned}
\label{L_detco}
&\mathcal{L}\left(\boldsymbol{I}_q, \boldsymbol{I}_k, \boldsymbol{P}_q, \boldsymbol{P}_k\right)=\sum_{i=1}^4 w_i \cdot\left(\mathcal{L}_{g \leftrightarrow g}^i+\mathcal{L}_{l \leftrightarrow l}^i+\mathcal{L}_{g \leftrightarrow l}^i\right), 
\end{aligned}
\end{equation}
\begin{equation}
\begin{aligned}
\label{Lgg}
&\mathcal{L}_{g \leftrightarrow g}\left(\boldsymbol{I}_q, \boldsymbol{I}_k\right)=-\log \frac{\exp \left(q^g \cdot k_{+}^g / \tau\right)}{\sum_{i=0}^K \exp \left(q^g \cdot k_i^g / \tau\right)},
\end{aligned}
\end{equation}
\begin{equation}
\begin{aligned}
\label{Lll}
&\mathcal{L}_{l \leftrightarrow l}\left(\boldsymbol{P}_q, \boldsymbol{P}_k\right)=-\log \frac{\exp \left(q^l \cdot k_{+}^l / \tau\right)}{\sum_{i=0}^K  \exp \left(q^l \cdot k_i^l / \tau\right)}, 
\end{aligned}
\end{equation}
\begin{equation}
\begin{aligned}
\label{Lgl}
&\mathcal{L}_{g \leftrightarrow l}\left(\boldsymbol{P}_q, \boldsymbol{I}_k\right)=-\log \frac{\exp \left(q^l \cdot k_{+}^g / \tau\right)}{\sum_{i=0}^K \exp \left(q^l \cdot k_i^g / \tau\right)},
\end{aligned}
\end{equation}
where Eq.~\ref{L_detco} is a multistage contrastive loss, Eq.~\ref{Lgg} to Eq.~\ref{Lgl} are global losses, local loss and local and global cross loss, respectively.  $q$ refers to query image, $k$ refers to key images, $k_{+}$ is a single key in the dictionary that $q$ matches, $w_i$ refers to the loss weight in the $i$-th level, $\tau$ is a temperature hyper-parameter.

\vspace{-2mm}
\subsection{B-line detector}
\label{sec:2.2}

\noindent \textbf{Faster R-CNN} \cite{ren2015faster} is a two-stage object detection network structure. The features extracted by the ResNet (see Section \ref{sec:2.1}) are also shared with the region proposal network, where anchors for predicted objectives are proposed. A number of anchors that have higher detection confidence score than the threshold will be selected by non-maximum suppression algorithm. 

\noindent \textbf{Loss functions}:
For  fine-tuning the Faster R-CNN, the total loss is a combination of classification loss and localization loss as described in Eq.~\ref{L_total}, 
\begin{equation}
\label{L_total}
\mathcal{L}_{\text {total}}=\left(\lambda_{\text {RPN}}\cdot\mathcal{L}_{\text {RPN}}\right)+\left(\lambda_{\text {FastRCNN}}\cdot\mathcal{L}_{\text {FastRCNN}}\right)
\end{equation}
where $\lambda_{\text {RPN}}$ and $\lambda_{\text {FastRCNN}}$ are the weights for region proposal layers \cite{ren2015faster} and Fast-RCNN \cite{girshick2015fast}. Both are usually set to 1. $\mathcal{L}_{\text{RPN}}$ and $\mathcal{L}_{\text{FastRCNN}}$ are defined in Eq.~\ref{L_rpn} to Eq.~\ref{L_eiou}.
\begin{equation}
\begin{aligned}
\label{L_rpn}
&\mathcal{L}_{\text{RPN}}=\mathcal{L}_{cls}+\mathcal{L}_{\text {detect}}, %EIOU}},
\end{aligned}
\end{equation}
\begin{equation}
\begin{aligned}
\label{L_fast}
\mathcal{L}_{FastRCNN}=-\log p_i + \mathcal{L}_{\text {detect}},%EIOU}},
\end{aligned}
\end{equation}
\begin{equation}
\begin{aligned}
\label{L_cls}
&\mathcal{L}_{cls}=-p_i^* \log p_i-\left(1-p_i^*\right) \log \left(1-p_i\right),
\end{aligned}
\end{equation}
where $\mathcal{L}_{cls}$ is a cross-entropy loss in the region proposal layer. In this paper, the regions are classified into background or B-lines. $\mathcal{L}_{\text {detect}}$ refers to the loss in bounding box (BB) regression  $p_i^*$ is a binary ground truth label specifying whether anchor $i$ is a B-line. $p_i$ is the predicted probability of anchor $i$ being a B-line. 

The original Faster-RCNN employs smooth-$l_1$ loss as   $\mathcal{L}_{\text {detect}}$. However, our aim is to detect long B-lines, so we propose using an extension of Intersection over Union-based loss, called an efficient IoU-based loss  $\mathcal{L}_{\text {EIOU}}$ \cite{zhang2022focal}, defined as follows. 
\begin{equation}
\begin{aligned}
\label{L_eiou}
\mathcal{L}_{\text {IOU}} &= 1-\frac{|B \cap B^{gt}|}{|B \cup B^{gt}|} \\
\mathcal{L}_{\text {EIOU}} &= \mathcal{L}_{\text {IOU}}
+\frac{\rho^2\left(\mathbf{b}, \mathbf{b}^{\mathbf{g t}}\right)}{\left(w^c\right)^2+\left(h^c\right)^2}
+\frac{\rho^2\left(w, w^{g t}\right)}{\left(w^c\right)^2}+\frac{\rho^2\left(h, h^{g t}\right)}{\left(h^c\right)^2},
\end{aligned}
\end{equation}
where $B$ and $B^{gt}$ are the predicted box and the ground truth box area. $\mathbf{b}$ and $\mathbf{b^{gt}}$ denote the central points of $B$ and $B^{gt}$ respectively. $\rho^2\left(\cdot\right)$ indicates the Euclidean distance. $w^{c}$ and $h^{c}$ are the width and height of the smallest enclosing box covering the two boxes. $\mathcal{L}_{\text {EIOU}}$ explicitly measures the discrepancies of the overlap area, the central point, and the side length in the BB regression.\vspace{-3mm}

\section{Implementation Details}
\label{sec:3}

\begin{itemize}[leftmargin=0em]

\item[]	\textbf{Dataset}:  30 patients (age range 22 - 91 years old) were recruited from Paediatric Nephrology Unit, Meyer Childrens Hospital, Florence, Italy. For each patient, we applied the 8-site B-lines score technique \cite{Torino2020can} to collect LUS videos and/or images, ending up with 232 video series (total 23756 frames) and 246 images. The files were saved in DICOM format. We used MicroDicom$\footnote{https://www.microdicom.com/}$ to extract all the video frames to images, and then cropped the size to 616$\times$480 (width $\times$ height) to make sure the lung area is fully included.

\item[]	%\textbf{Dataset Setup}:
The LUS images used in this study consists of two parts: i) for unsupervised feature learning, we incorporated our data from 10 patients with the POCUS dataset\footnote{https://github.com/jannisborn/covid19\_ultrasound} to increase diversity of ultrasonic lung features. POCUS is an expanding ultrasound dataset that contains lung artefacts and healthy lung samples, and currently has more than 200 LUS videos and 59 images from both convex probes and linear probes. Together with 10 videos from the 10 patients, the final combined dataset contains 4407 unlabelled images in total after merging the videos to images. ii) For fine-tuning, the data from the remaining 20 patients serves the purpose of B-line detection. Since the contrastive learning method inherently does not require a large amount of labelled data, we selected 84 images and annotated the B-lines using Pair\footnote{https://www.aipair.com.cn/en/}providing the BB and the segmentation mask of B-lines. In the labelled dataset, 70\% of the images are used for training, and 30\% are used for evaluation. The output of the detection is in the  COCO format\footnote{https://cocodataset.org/\#home}. 

\item[]	\textbf{Unsupervised Representation Learning}:
The simulations were performed using the High Performance Computing (HPC) facility of the  University of Bristol. %, and repositories were installed referring to the requirements$\footnote{https://github.com/shuuchen/DetCo.pytorch/blob/main/requirements.txt}$ for unsupervised training. 
The network comprises a Resnet50 backbone, four global MLP heads and a local patch MLP head \cite{xie2021detco}. The training was done on 4 NVIDIA RTX2080Ti GPUs, using stochastic gradient descent with the initial learning rate of 0.015 for 200 epochs. All models were trained with a batch size of 4, which was the maximum allowed by the HPC memory. The hyper-parameters settings followed DetCo. Among the 200 epochs, model weights with the highest accuracy were saved.

\item[]	\textbf{Fine-tuning}: 
The weights of the best Resnet50 \cite{he2016deep} acquired from the first stage were loaded to a Faster R-CNN \cite{ren2015faster} for the B-line detection task. 
As suggested in the original DetCo paper~\cite{xie2021detco}, detection tasks were carried out based on detectron2$\footnote{https://github.com/facebookresearch/detectron2}$. Following the requirements, the network was trained for 1x schedule on Linux with Python = 3.8, Pytorch = 1.9.0 and torchvision = 0.10.0 on 2 NVIDIA Tesla P100 GPUs.

\end{itemize}
\vspace{0mm}

\section{Experimental Results}
\label{sec:4}

We tested our proposed framework (pre-trained on LUS image dataset and fine tuned with EIoU loss) with 25 LUS images. We compared our results with those of the models: i) pre-trained on ImageNet and fine tuned with smooth-$l_1$ loss, ii) pre-trained on LUS image dataset and fine tuned with smooth-$l_1$ loss, iii) pre-trained on LUS image dataset and fine tuned with IoU loss \cite{yu2016unitbox}, and iv) the model-based method, proposed in \cite{anantrasirichai2017line}. 
We set a detection threshold to 0.5 to compute the number of true positives, false positives and false negatives B-lines. Table \ref{tab:results} shows the performances in term of precision, recall, accuracy and F$_1$-scores.

By examining the metrics mentioned above, we can see that data-driven methods surpass model-based method. Amongst learning approaches, the proposed method achieves the highest recall (91.43\%), accuracy (84.21\%) and F$_1$-score (91.43\%). Whereas the method pretrained with ImageNet using smooth-$l_1$ loss performs the best in terms of precision, only a percentage of the detected B-lines are the actual B-lines. However, in the context of medical diagnostic, the miss detection is usually more undesirable, as it may lead to sever conditions being trivialized. Therefore, along with a high precision, the recall, the percentage of the actual B-lines detected, needs to be as high as possible. The results of our proposed method demonstrates the benefit of unsupervised training for learning features of the LUS images. Using IoU loss and EIoU loss in fine-tuning stage further increases the performance in terms of recall. The F$_1$ score is a measurement, which balances the precision and the recall. Our method achieves the highest F$_1$ score indicating the superior overall performance in B-line detection.

Fig. 3 shows some examples of the detected B-lines. It is obvious that the method in  \cite{anantrasirichai2017line} relies heavily on localising the detected pleural line, which may result in a wrong position or in classing a vertical artifact as a B-line. In contrast, our method detects B-lines without the need for localising pleural lines and has the ability to distinguish the true B-lines from the bright vertical lines that only look similar to B-lines. Nevertheless, for some obscure B-lines, our method still presents some miss detections or false positives.

\begin{figure}[htbp]
\label{compare}
\centering
\includegraphics[width=1\linewidth]{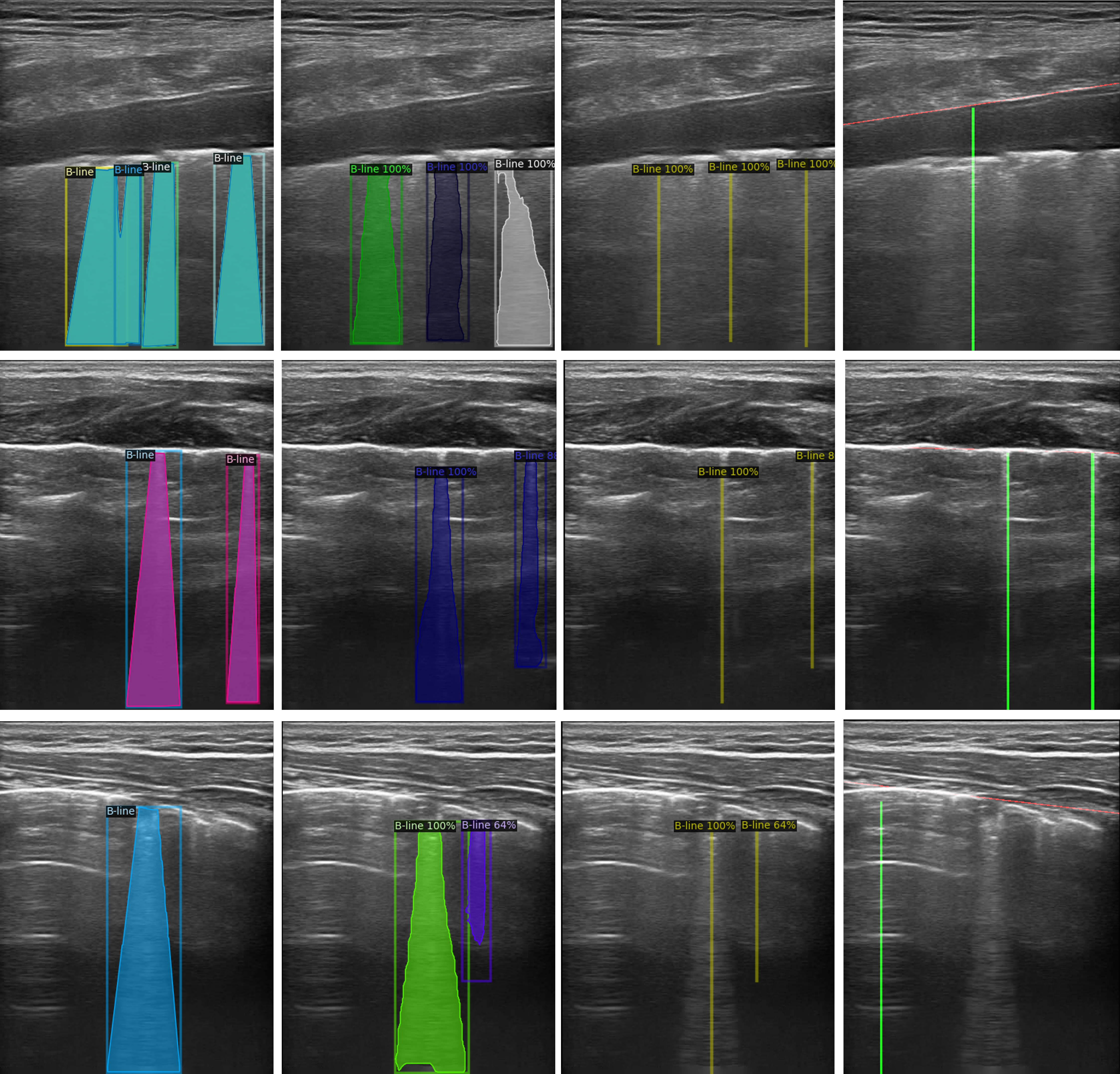}
\caption{B-line detection results. From left to right, each column represents ground truth, B-line detection results of the proposed method in segmentation form, B-lines detection results of the proposed method shown as center line and the model-based method \cite{anantrasirichai2017line} respectively.}
\vspace{-5mm}
\end{figure}

\begin{table}[htbp]
\caption{Detection results with various training procedures }\label{tab:results}
\resizebox{\columnwidth}{!}{%
\begin{tabular}{c|cccc}
\hline
                        & PRECISION         & RECALL           & ACCURACY         & F$_1$-score               \\ \hline
ImageNet+smooth-$l_1$       & \textbf{100.00\%} & 80.65\%          & 80.65\%          & 89.29\%          \\
LUS+smooth-$l_1$            & 96.67\%           & 82.86\%          & 80.56\%          & 89.23\%          \\
LUS+IoU                 & 90.00\%           & 84.38\%          & 77.14\%          & 87.10\%          \\
%LUS+DIoU                & 93.33\%           & 80.00\%          & 75.68\%          & 86.15\%          \\
LUS+EIoU (proposed)               & 91.43\%           & \textbf{91.43\%} & \textbf{84.21\%} & \textbf{91.43\%} \\
model-based \cite{anantrasirichai2017line} & 55.56\%           & 42.86\%          & 31.91\%          & 48.39\%          \\ \hline
\end{tabular}%
}
\end{table}
\vspace{-5mm}  

\section{Conclusions}
\label{sec:4}
\vspace{-2mm}
This study demonstrates the superiority of contrastive learning for B-line detection in LUS images and investigates the performance of three loss functions. Without the need for large amounts of ground truth data or the need for pleural line localisation, the proposed method's performance surpasses that of model-based approaches. It is also shown to be the most suitable for B-line detection in clinical applications, among all tested approaches. The limitation of this study is that the data was annotated by only one expert, and this may bring bias, so more validation needs to be done in future work.

\small
\bibliographystyle{IEEEtran}
\bibliography{LUSDetCoRef.bib}

\begin{thebibliography}{10}
\providecommand{\url}[1]{#1}
\csname url@samestyle\endcsname
\providecommand{\newblock}{\relax}
\providecommand{\bibinfo}[2]{#2}
\providecommand{\BIBentrySTDinterwordspacing}{\spaceskip=0pt\relax}
\providecommand{\BIBentryALTinterwordstretchfactor}{4}
\providecommand{\BIBentryALTinterwordspacing}{\spaceskip=\fontdimen2\font plus
\BIBentryALTinterwordstretchfactor\fontdimen3\font minus
  \fontdimen4\font\relax}
\providecommand{\BIBforeignlanguage}[2]{{%
\expandafter\ifx\csname l@#1\endcsname\relax
\typeout{** WARNING: IEEEtran.bst: No hyphenation pattern has been}%
\typeout{** loaded for the language `#1'. Using the pattern for}%
\typeout{** the default language instead.}%
\else
\language=\csname l@#1\endcsname
\fi
#2}}
\providecommand{\BIBdecl}{\relax}
\BIBdecl

\bibitem{touw2015lung}
H.~Touw, P.~Tuinman, H.~Gelissen, E.~Lust, and P.~Elbers, ``Lung ultrasound:
  routine practice for the next generation of internists,'' \emph{Neth J Med},
  vol.~73, no.~3, pp. 100--107, 2015.

\bibitem{touw2019routine}
H.~Touw, A.~E. Schuitemaker, F.~Daams, D.~L. van~der Peet, E.~Bronkhorst,
  P.~Schober, C.~Boer, and P.~R. Tuinman, ``Routine lung ultrasound to detect
  postoperative pulmonary complications following major abdominal surgery: a
  prospective observational feasibility study,'' \emph{The Ultrasound Journal},
  vol.~11, no.~1, pp. 1--8, 2019.

\bibitem{hecking2015consequences}
M.~Hecking, H.~Rayner, and P.~Wabel, ``What are the consequences of volume
  expansion in chronic dialysis patients? defining and measuring fluid overload
  in hemodialysis patients,'' in \emph{Seminars in dialysis}, vol.~28,
  no.~3.\hskip 1em plus 0.5em minus 0.4em\relax Wiley Online Library, 2015, pp.
  242--247.

\bibitem{torino2016agreement}
C.~Torino, L.~Gargani, R.~Sicari, K.~Letachowicz, R.~Ekart, D.~Fliser,
  A.~Covic, K.~Siamopoulos, A.~Stavroulopoulos, Z.~A. Massy \emph{et~al.},
  ``The agreement between auscultation and lung ultrasound in hemodialysis
  patients: the lust study,'' \emph{Clinical Journal of the American Society of
  Nephrology}, vol.~11, no.~11, pp. 2005--2011, 2016.

\bibitem{vitturi2014lung}
N.~Vitturi, M.~Dugo, M.~Soattin, F.~Simoni, L.~Maresca, R.~Zagatti, and M.~C.
  Maresca, ``Lung ultrasound during hemodialysis: the role in the assessment of
  volume status,'' \emph{International urology and nephrology}, vol.~46, no.~1,
  pp. 169--174, 2014.

\bibitem{panuccio2012chest}
V.~Panuccio, G.~Enia, R.~Tripepi, C.~Torino, M.~Garozzo, G.~G. Battaglia,
  C.~Marcantoni, L.~Infantone, G.~Giordano, M.~L. De~Giorgi \emph{et~al.},
  ``Chest ultrasound and hidden lung congestion in peritoneal dialysis
  patients,'' \emph{Nephrology Dialysis Transplantation}, vol.~27, no.~9, pp.
  3601--3605, 2012.

\bibitem{lichtenstein1997comet}
D.~Lichtenstein, G.~Meziere, P.~Biderman, A.~Gepner, and O.~Barre, ``The
  comet-tail artifact: an ultrasound sign of alveolar-interstitial syndrome,''
  \emph{American journal of respiratory and critical care medicine}, vol. 156,
  no.~5, pp. 1640--1646, 1997.

\bibitem{noble2009ultrasound}
V.~E. Noble, A.~F. Murray, R.~Capp, M.~H. Sylvia-Reardon, D.~J. Steele, and
  A.~Liteplo, ``Ultrasound assessment for extravascular lung water in patients
  undergoing hemodialysis: time course for resolution,'' \emph{Chest}, vol.
  135, no.~6, pp. 1433--1439, 2009.

\bibitem{fu2021lung}
Q.~Fu, Z.~Chen, J.~Fan, C.~Ling, X.~Wang, X.~Liu, and Y.~Shen, ``Lung
  ultrasound methods for assessing fluid volume change and monitoring dry
  weight in pediatric hemodialysis patients,'' \emph{Pediatric Nephrology},
  vol.~36, no.~4, pp. 969--976, 2021.

\bibitem{allinovi2017lung}
M.~Allinovi, M.~Saleem, P.~Romagnani, P.~Nazerian, and W.~Hayes, ``Lung
  ultrasound: a novel technique for detecting fluid overload in children on
  dialysis,'' \emph{Nephrology Dialysis Transplantation}, vol.~32, no.~3, pp.
  541--547, 2017.

\bibitem{soldati2009sonographic}
G.~Soldati, R.~Copetti, and S.~Sher, ``Sonographic interstitial syndrome: the
  sound of lung water,'' \emph{Journal of Ultrasound in Medicine}, vol.~28,
  no.~2, pp. 163--174, 2009.

\bibitem{anantrasirichai2016automatic}
N.~Anantrasirichai, M.~Allinovi, W.~Hayes, and A.~Achim, ``Automatic b-line
  detection in paediatric lung ultrasound,'' in \emph{2016 IEEE International
  Ultrasonics Symposium (IUS)}.\hskip 1em plus 0.5em minus 0.4em\relax IEEE,
  2016, pp. 1--4.

\bibitem{brattain2013automated}
L.~J. Brattain, B.~A. Telfer, A.~S. Liteplo, and V.~E. Noble, ``Automated
  b-line scoring on thoracic sonography,'' \emph{Journal of Ultrasound in
  Medicine}, vol.~32, no.~12, pp. 2185--2190, 2013.

\bibitem{anantrasirichai2017line}
N.~Anantrasirichai, W.~Hayes, M.~Allinovi, D.~Bull, and A.~Achim, ``Line
  detection as an inverse problem: application to lung ultrasound imaging,''
  \emph{IEEE transactions on medical imaging}, vol.~36, no.~10, pp. 2045--2056,
  2017.

\bibitem{moshavegh2018automatic}
R.~Moshavegh, K.~L. Hansen, H.~M{\o}ller-S{\o}rensen, M.~B. Nielsen, and J.~A.
  Jensen, ``Automatic detection of b-lines in $ in vivo $ lung ultrasound,''
  \emph{IEEE transactions on ultrasonics, ferroelectrics, and frequency
  control}, vol.~66, no.~2, pp. 309--317, 2018.

\bibitem{karakucs2020detection}
O.~Karaku{\c{s}}, N.~Anantrasirichai, A.~Aguersif, S.~Silva, A.~Basarab, and
  A.~Achim, ``Detection of line artifacts in lung ultrasound images of covid-19
  patients via nonconvex regularization,'' \emph{IEEE transactions on
  ultrasonics, ferroelectrics, and frequency control}, vol.~67, no.~11, pp.
  2218--2229, 2020.

\bibitem{karakucs2020convergence}
O.~Karaku{\c{s}}, P.~Mayo, and A.~Achim, ``Convergence guarantees for
  non-convex optimisation with cauchy-based penalties,'' \emph{IEEE
  Transactions on Signal Processing}, vol.~68, pp. 6159--6170, 2020.

\bibitem{van2019localizing}
R.~J. Van~Sloun and L.~Demi, ``Localizing b-lines in lung ultrasonography by
  weakly supervised deep learning, in-vivo results,'' \emph{IEEE journal of
  biomedical and health informatics}, vol.~24, no.~4, pp. 957--964, 2019.

\bibitem{selvaraju2017grad}
R.~R. Selvaraju, M.~Cogswell, A.~Das, R.~Vedantam, D.~Parikh, and D.~Batra,
  ``Grad-cam: Visual explanations from deep networks via gradient-based
  localization,'' in \emph{Proceedings of the IEEE international conference on
  computer vision}, 2017, pp. 618--626.

\bibitem{Baloescu2020automated}
C.~Baloescu, G.~Toporek, S.~Kim, K.~McNamara, R.~Liu, M.~M. Shaw, R.~L.
  McNamara, B.~I. Raju, and C.~L. Moore, ``Automated lung ultrasound b-line
  assessment using a deep learning algorithm,'' \emph{IEEE Transactions on
  Ultrasonics, Ferroelectrics, and Frequency Control}, vol.~67, no.~11, pp.
  2312--2320, 2020.

\bibitem{Kerdegari2021bline}
\BIBentryALTinterwordspacing
H.~Kerdegari, N.~T.~H. Phung, A.~McBride, L.~Pisani, H.~V. Nguyen, T.~B. Duong,
  R.~Razavi, L.~Thwaites, S.~Yacoub, A.~Gomez, and V.~Consortium, ``B-line
  detection and localization in lung ultrasound videos using spatiotemporal
  attention,'' \emph{Applied Sciences}, vol.~11, no.~24, 2021. [Online].
  Available: \url{https://www.mdpi.com/2076-3417/11/24/11697}
\BIBentrySTDinterwordspacing

\bibitem{hadsell2006dimensionality}
R.~Hadsell, S.~Chopra, and Y.~LeCun, ``Dimensionality reduction by learning an
  invariant mapping,'' in \emph{2006 IEEE Computer Society Conference on
  Computer Vision and Pattern Recognition (CVPR'06)}, vol.~2.\hskip 1em plus
  0.5em minus 0.4em\relax IEEE, 2006, pp. 1735--1742.

\bibitem{xie2021detco}
E.~Xie, J.~Ding, W.~Wang, X.~Zhan, H.~Xu, P.~Sun, Z.~Li, and P.~Luo, ``Detco:
  Unsupervised contrastive learning for object detection,'' in
  \emph{Proceedings of the IEEE/CVF International Conference on Computer
  Vision}, 2021, pp. 8392--8401.

\bibitem{he2016deep}
K.~He, X.~Zhang, S.~Ren, and J.~Sun, ``Deep residual learning for image
  recognition,'' in \emph{Proceedings of the IEEE conference on computer vision
  and pattern recognition}, 2016, pp. 770--778.

\bibitem{ren2015faster}
S.~Ren, K.~He, R.~Girshick, and J.~Sun, ``Faster r-cnn: Towards real-time
  object detection with region proposal networks,'' \emph{Advances in neural
  information processing systems}, vol.~28, 2015.

\bibitem{he2020momentum}
K.~He, H.~Fan, Y.~Wu, S.~Xie, and R.~Girshick, ``Momentum contrast for
  unsupervised visual representation learning,'' in \emph{Proceedings of the
  IEEE/CVF conference on computer vision and pattern recognition}, 2020, pp.
  9729--9738.

\bibitem{yang2021instance}
C.~Yang, Z.~Wu, B.~Zhou, and S.~Lin, ``Instance localization for
  self-supervised detection pretraining,'' in \emph{Proceedings of the IEEE/CVF
  Conference on Computer Vision and Pattern Recognition}, 2021, pp. 3987--3996.

\bibitem{girshick2015fast}
R.~Girshick, ``Fast r-cnn,'' in \emph{Proceedings of the IEEE international
  conference on computer vision}, 2015, pp. 1440--1448.

\bibitem{zhang2022focal}
Y.-F. Zhang, W.~Ren, Z.~Zhang, Z.~Jia, L.~Wang, and T.~Tan, ``Focal and
  efficient iou loss for accurate bounding box regression,''
  \emph{Neurocomputing}, vol. 506, pp. 146--157, 2022.

\bibitem{Torino2020can}
\BIBentryALTinterwordspacing
C.~Torino, R.~Tripepi, C.~Loutradis, P.~Sarafidis, G.~Tripepi, F.~Mallamaci,
  and C.~Zoccali, ``Can the assessment of ultrasound lung water in
  haemodialysis patients be simplified?'' \emph{Nephrology Dialysis
  Transplantation}, vol.~36, no.~12, pp. 2321--2326, 12 2020. [Online].
  Available: \url{https://doi.org/10.1093/ndt/gfaa285}
\BIBentrySTDinterwordspacing

\bibitem{yu2016unitbox}
J.~Yu, Y.~Jiang, Z.~Wang, Z.~Cao, and T.~Huang, ``Unitbox: An advanced object
  detection network,'' in \emph{Proceedings of the 24th ACM international
  conference on Multimedia}, 2016, pp. 516--520.

\end{thebibliography}

\end{document}